\documentstyle[12pt]{article}

\advance\textheight by \topskip
\newcommand{\sla}{\kern -5.4pt /}
\newcommand{\slaar}{\kern -7. pt / \kern 3.pt}
\newcommand{\Dir}{\kern -6.4pt\Big{/}}%su lettere italiane minuscole
\newcommand{\Dirin}{\kern -10.4pt\Big{/}\kern 4.4pt}
\newcommand{\DDir}{\kern -7.6pt\Big{/}}%su lettere italiane maiuscole
\newcommand{\DGir}{\kern -6.0pt\Big{/}}%su lettere greche

\newcommand{\ra}{\rightarrow}
\newcommand{\be}{\begin{equation}}
\newcommand{\ee}{\end{equation}}
\newcommand{\bea}{\begin{eqnarray}}
\newcommand{\eea}{\end{eqnarray}}
\newcommand{\beanon}{\begin{eqnarray*}}
\newcommand{\eeanon}{\end{eqnarray*}}
\newcommand{\ba}{\begin{array}}
\newcommand{\ea}{\end{array}}
\newcommand{\bd}{\begin{description}}
\newcommand{\ed}{\end{description}}
\newcommand{\bi}{\begin{itemize}}
\newcommand{\ei}{\end{itemize}}
\newcommand{\ben}{\begin{enumerate}}
\newcommand{\een}{\end{enumerate}}
\newcommand{\bc}{\begin{center}}
\newcommand{\ec}{\end{center}}

\newcommand{\ol}{\overline}

\newcommand{\ar}{\rightarrow}

\newcommand{\hsk}{\hskip 10 pt\noindent}

\def\wm{M_{_W}}
\def\zm{M_{_Z}}
\def\gf{G_{\mu}}
\def\pd{{\it production $\times$ decay\ }}
\def\epem{$e^+ e^-$\ }
\def\bb{$b \bar b$\ }
\def\app #1 #2 #3 {{\it  Acta Phys.Polon.} {#1} (#2) #3\ }
\def\ap #1 #2 #3 {{\it Ann. Phys. }{ #1} (#2) #3\ }
\def\intj #1 #2 #3{{\it Int. J. Mod. Phys.} {#1} (#2) #3\ }
\def\hpa #1 #2 #3{{\it Helv. Phys. Acta. }{ #1} #2) #3\ }
\def\mpl #1 #2 #3 {{\it Mod.~Phys.~Lett.} {#1} (#2) #3\ }
\def\np #1 #2 #3 {{\it Nucl.~Phys.} {#1} (#2) #3\ }
\def\pl #1 #2 #3 {{\it Phys.~Lett.} {#1} (#2) #3\ }
\def\pr #1 #2 #3 {{\it Phys.~Rev.} {#1} (#2) #3\ }
\def\prep #1 #2 #3 {{\it Phys.~Rep.} {#1} (#2) #3\ }
\def\prl #1 #2 #3 {{\it Phys.~Rev.~Lett.} {#1} (#2) #3\ }
\def\rmp #1 #2 #3 {{\it Rev. Mod. Phys.} {#1} (#2) #3\ }
\def\zp #1 #2 #3 {{\it Z.~Phys.} {#1} (#2) #3\ }
\def\cpc #1 #2 #3 {{\it Comp. Phys. Commun.} {#1} (#2) #3\ }
\def\xx #1 #2 #3 {{#1}, (#2) #3\ }

\begin{document}
\tolerance=100000
\thispagestyle{empty}
\setcounter{page}{0}

\begin{flushright}
{\large DFTT 52/97}\\
{\rm August 1997\hspace*{.5 truecm}}\\
\end{flushright}

\vspace*{\fill}

{\Large \bf \noindent
Six fermion processes at future $e^+ e^-$ colliders: signal
and irreducible background for top, WWZ and Higgs  physics in charged current
 final states.
\footnote{ Work supported in part by Ministero
dell' Universit\`a e della Ricerca Scientifica.\\[2 mm]
e-mail: accomando,ballestrero,pizzio@to.infn.it}}\\[2.cm]
\bc
{\large Elena Accomando, Alessandro Ballestrero and Marco Pizzio}\\[.3 cm]
{\it I.N.F.N., Sezione di Torino, Italy}\\
{\it and}\\
{\it Dipartimento di Fisica Teorica, Universit\`a di Torino, Italy}\\
{\it v. Giuria 1, 10125 Torino, Italy.}\\
\ec

\vspace*{\fill}

\begin{abstract}
{\normalsize\noindent
 We compute several total and differential cross sections relevant to top,
 WWZ and Higgs physics at future $e^+e^-$ colliders taking into account the 
full set of Feynman diagrams for six fermion final states. 
We examine in particular charged current processes,   in which
 final particles cannot be formed by three Z's decay.
We  include in our calculations initial state radiation and beamstrahlung 
effects, and the most important QCD corrections in an approximate 
(naive) form.  We also compare this complete approach with  \pd approximation.
} 
\end{abstract}
\vspace*{\fill}
{\it Contribution to the Proceedings of the "ECFA/DESY Study on
  Physics and Detectors for the Linear Collider",  February to November 1996,
  ed. R. Settles, Desy 97-123E. }
%.  
\newpage

\section{ Introduction }
Processes with many particles in the final state become more and more important
at high energies for present and future accelerators. At LEP2 WW
and Higgs physics has lead to a careful study of all complete four fermion
final states, for which electroweak dedicated codes have been constructed
(for a review on the argument see \cite{yr}\cite{wweg}\cite{cpc}).
When future $e^+ e^-$ colliders will come into operation, 
states with even more final particles will become important.
In particular all physics regarding  top studies, three vector boson production
and intermediate Higgs searches  will deal  with six fermion final states.

The complete calculation of such  processes is rather complicated even at tree
level: it requires the computation of hundreds of Feynman diagrams. 
The alternative is that of considering only the most important contributions
to them, which generally correspond to  diagrams in which 
unstable particles (W, Z, top, Higgs) can go on mass shell, and evaluate
them in the \pd approximation. With such a method it is also easier to
compute  the most important higher order corrections. On the other hand
one is neglecting the contributions of a lot of diagrams, interference effects,
spin correlations effects, off-shellness of the resonant diagrams, and only 
the complete calculation can establish
case by case the reliability of such approximations, which often depends on 
final states, cuts, energies and  distributions.

All processes $e^+ e^- \ra six fermions$ can be divided
according to their final states. As for four fermion final states, we can 
identify Charged Currents (CC), Neutral Currents (NC) and 
Mixed processes (MIX) \cite{class}\cite{wweg}\cite{cpc}. 
We   call CC the six fermion processes in which
the final particles  can form two W's and a Z but not three Z's (e. g. $\mu
\bar \nu u \bar d e^+ e^- $),
NC those in which three Z's and not two W's and a Z can be formed 
(e. g. $u \bar u \mu^+  \mu^- e^+ e^-$), MIX
those in which  both can be formed (e. g. $u \bar u d \bar d e^+ e^-$).

As already mentioned, these reactions are of particular interest because they 
are related
to top, WWZ and Higgs physics. If we consider for instance a final state
like $b \bar b \mu \bar \nu_\mu u \bar d$, this may be produced  by the decay
of WWZ or come from two W's and two b's, which in turn may descend from
two tops. WWZ intermediate state may be due to Higgs-Z production with
an Higgs decaying into two W's.  
 Of course the various channels are not separated
in the reality, and many more diagrams (irreducible background) contribute to
such processes. They can be distinguished only for their different contribution
to various zones of the phase space and can eventually be disentangled by 
applying experimental cuts. 
In order to study such cuts, and to evaluate the magnitude
of the various contributions after they have been applied, one must
use the full calculation.

We have produced a code (SIXPHACT) to compute all six fermions CC processes.
With it we will examine top top in the continuum, WWZ and intermediate
Higgs events. In all three cases the signal intermediate state contains
two W's. We will in particular discuss those processes in which (at least) one
isolated lepton (e.g. a $\mu^-$ without the corresponding $\mu^+$) 
indicates the presence of two W's. A part of this analysis has already been
reported in ref~\cite{noi} to which we refer for details. 
 Top  and Higgs physics have  already been 
analyzed within a four particles ($b \bar b W^+W^-$) final 
state  approximation \cite{bbww}.  Six fermion physics has also been considered
in \cite{pv}\cite{kuri}. In particular in ref.\cite{pv} the processes
$e^+e^-\ar \mu^+\mu^-\tau^- \bar\nu_\tau u\bar d$ and 
$e^+e^-\ar \mu^+\mu^- e^- \bar\nu_e u\bar d$ have been analyzed for their
interest in  Higgs searches, while ref.\cite{kuri} deals with
the reaction $e^+e^-\ar b \bar b u\bar d \mu^- \bar \nu_\mu$ and its 
relation to top physics.

To give an idea of the complexity of the problem, we remind that the process 
$e^+e^-\ar b \bar b u\bar d \mu^- \bar \nu_\mu$ has 232 tree level diagrams, 
$e^+e^-\ar e^- \bar \nu_e u \bar d s \bar s$ 420, and 
$e^+e^-\ar e^- \bar \nu_e u \bar d e^- e^+$ 1254. The integration variables
are  at least thirteen, but they   become  seventeen 
for the more realistic case in which  initial state radiation (ISR) and 
beamstrahlung (BST) are accounted for. For such reasons it is extremely 
important to use a method for computing helicity 
amplitudes which allows a very fast and precise computation.  
We have used to this end  PHACT \cite{phact}, a set of routines based on
the method of ref.\cite{method}, in order to generate the fortran code
for the amplitudes. We have computed all Feynman diagrams by calculating 
subdiagrams  of increasing complexity and reusing them whenever needed. 
The method used is particularly suited for this procedure.
We have introduced ISR via the structure function method \cite{sf} and
BST  with a link to the program CIRCE~\cite{circe}. Different
phase spaces with different mappings to account for various peak structures
have been employed. The numerical integration has been performed with 
VEGAS\cite{vegas}.

As far as QCD corrections are concerned, we have introduced them in the so
called naive QCD approach (NQCD). This amounts to consider that we have
diagrams with vertices which in narrow width approximation (NWA) correspond to 
\pd of W's, Z's, t's, h's. The corresponding corrections for the decay  are 
factorized and we multiply such diagrams for these factors.
We have also included QCD corrections in the 
naive formulation to the $t\bar t$ production vertices. The first order QCD 
corrections~\cite{Zttcor} to $\sigma_{VV}$ and $\sigma_{AA}$, the vector-vector
and axial-axial contributions to  the total on-shell top top 
cross section, factorize separately. We have introduced  these corrections,
but we have not applied any 
correction to the interference term $\sigma_{VA}$ which vanishes when
integrated over the full phase space. Our treatment of QCD corrections is in
any case exact only in the narrow width approximation 
 for the total cross section with no cuts.
In all other cases it must be considered as a rough estimate of the
most important QCD corrections. In many
cases this has however proved to be a reasonable approximation,
probably just because the error
on the corrections  reflects upon a much smaller error on the cross sections.

For the numerical part we have used the $\rm G_{\mu}$-scheme 
\begin{eqnarray}
s_{_W}^2 = 1 - {{\wm^2}\over {\zm^2}}, \qquad 
g^2 &=& 4{\sqrt 2}\gf\wm^2
\end{eqnarray}
and the  input masses  $m_Z=91.1888$ GeV, $m_W=80.23$ GeV. We have chosen
 $m_t=180$ GeV and for $m_b$ the running mass value $m_b=2.7$ GeV. 
For the strong corrections to  Z, W  top and Higgs decay widths
and vertices  we have used
$\alpha_s(m_Z)=$ 0.123 and evoluted it to the appropriate scales.

We have moreover implemented the following general set of cuts :
\bc
 jet(quark) energy $> 3$ GeV;\quad
 lepton energy $> 1$ GeV;\quad
 jet-jet mass $> 10$~GeV;\\
 lepton-beam angle $> 10^{\circ}$;\quad
 jet-beam angle $> 5^{\circ}$; \quad
 lepton-jet angle $> 5^{\circ}$.
\ec

Other cuts specific to particular studies will be described in the following.

\section {Top in the continuum}

After top discovery  and the  measurement of its mass (m$_t$= 174 $\pm$ 6 GeV)
 \cite{top}, its properties must be determined with high-precision.
 An extremely important characteristic of the top is  its
lifetime, which is much shorter than the time scale of strong interactions,
allowing to study it in the context of perturbative QCD. 
 The opening of a new channel and perturbative effects \cite{kuhn} determine  a 
sharp rise of the cross section at threshold. 
For such a reason, the best strategy  to measure
top  mass at future $e^+e^-$ colliders consists in running at $t\bar t$ 
threshold, while its static properties, such as magnetic and
electric dipole moments \cite{ttform} will be measured with high accuracy in 
the continuum, at higher energies. 
In any case,  future $e^+e^-$ colliders will produce a great number of top top
events: at 500 GeV the cross
section $\rm \sigma(t\bar t)$ is of the order of .5 pb, which corresponds
to about $2.5\times 10^4$ events  for an integrated luminosity of 50 $\rm
fb^{-1}$.
 
The  results in the following refer only to 
the continuum top top production. The tree level matrix elements for the 
complete calculation of the
various final states do not  take into account the
above mentioned threshold corrections. They can nevertheless be useful to
estimate   the  relevance of the irreducible
background due to all non double resonant diagrams also at threshold.

We  will consider two specific CC final states: 
$e^+e^-\ra \mu\bar\nu_{\mu} u\bar d
b \bar b$ and $e^+e^-\ra e \bar\nu_e u\bar d b \bar b$. 
The cross sections  due to  signal 
diagrams only, and the irreducible backgrounds due to 
all other diagrams (207 for the $\mu$ 
and 416 for the $e$ without Higgs)  and their interference with 
the signal at 500 and 800 GeV  are given in table~\ref{tabcstop}.
\begin{table}[hbt]\centering
\begin{tabular}{|c|c|c|c|} 
\hline
\rule [-0.25 cm]{0 cm}{0.75 cm}
$\sqrt{s}$ GeV& channel & $t\bar t$ signal (fb) & background (fb) \\ \hline
\hline
\rule [-.25 cm] {0 cm} {0.75 cm} 
500 & $\mu \bar \nu u\bar d b\bar b$ & 19.850(4) & 0.736(3) \\ \cline{2-2}
\cline{4-4}
\rule [-.25 cm] {0 cm} {0.75 cm} 
  & $e \bar \nu u\bar d b\bar b$  &  & 0.778(5) \\ \hline
\rule [-0.25 cm]{0 cm}{0.75 cm}
800 &$\mu \bar \nu u\bar d b\bar b$ & 10.700(2) & 1.007(4) \\ \cline{2-2}
\cline{4-4}
\rule [-0.25 cm]{0 cm}{0.75 cm}
 &$e \bar \nu u\bar d b\bar b$ &  & 1.21(2) \\ \hline
\end{tabular}
\caption{Cross section for the processes $e^+ e^-~\ra\;\mu\bar
\nu u\bar d b\bar b$ and $e^+ e^- \ra\; e\bar \nu u\bar d b\bar
b$.}
\label{tabcstop}
\end{table}
Here and in the following  we report between parenthesis 
the  statistical integration errors on the last digit of the result.

These values have been obtained taking into account b masses, the full set
of diagrams (without Higgs), ISR, BST, NQCD corrections for the decay vertices 
of the W's, the Z, tops and also for Z($\gamma$)tt vertex, as already explained.

\begin{table}[hbt]\centering
\begin{tabular}{|c|c|c|}
\hline
\rule [-0.25 cm]{0 cm}{0.75 cm}
$e^+ e^- \ra \mu\bar \nu u\bar d b\bar b$ & $t\bar t$ signal (fb)& background
(fb)\\ \hline \hline
\rule [-0.25 cm]{0 cm}{0.75 cm}
$\rm NWA$                        & 18.880(3) & 0.848(3)\\ \hline
\rule [-0.25 cm]{0 cm}{0.75 cm}
$\rm Born$                       & 18.286(3) & 0.824(3)\\ \hline
\rule [-0.25 cm]{0 cm}{0.75 cm}
$\rm ISR$                        & 17.419(3) & 0.750(3)\\ \hline
\rule [-0.25 cm]{0 cm}{0.75 cm}
$\rm ISR\; NQCD^*$                   & 17.188(3) & 0.753(3)\\ \hline
\rule [-0.25 cm]{0 cm}{0.75 cm}
$\rm ISR\; NQCD^*\; BST$               & 17.303(3) & 0.731(3)\\ \hline
\rule [-0.25 cm]{0 cm}{0.75 cm}
$\rm ISR\; NQCD^*\; SBAND$             & 17.308(3) & 0.728(3)\\ \hline
\rule [-0.25 cm]{0 cm}{0.75 cm}
$\rm ISR\; NQCD^*\; BST\; m_b$         & 17.352(3) & 0.736(3)\\ \hline
\rule [-0.25 cm]{0 cm}{0.75 cm}
$\rm ISR\; NQCD\; BST\; m_b$& 19.850(4) &  0.736(3)\\ \hline
\end{tabular}
\caption[] { Cross section for the process $e^+ e^- \ra\; \mu\bar
\nu u\bar d b\bar b$ at $\sqrt{s}=500$ GeV for different sets of
approximations}
\label{tabcstt500}
\end{table}

In table~\ref{tabcstt500}
we report the cross sections for $e^+e^-\ra \mu\bar\nu_{\mu} u\bar d b \bar b$
for different approximations, in order to understand the relevance
of these corrections.
The first result (NWA) reproduces what one would obtain using on shell 
calculation of $t\bar t$ production with subsequent decays of the tops to $bW$ 
and on shell decay of the $W$'s. 
The second result corresponds to the Born approximation, to which we add in the
following ones ISR, naive QCD corrections to $t$'s, $W$'s and $Z$ decay 
vertices (${\rm NQCD^*}$), these plus that of $Z(\gamma)t\bar t$ vertex (NQCD),
beamstrahlung for the TESLA parametrization (BST) and for the SBAND one.
The results of the last two lines have been obtained taking exactly into 
account  the mass  ($\rm m_b$) of the b's, which were considered massless 
in the previous ones in table~\ref{tabcstt500}.

\begin{figure}[h]
\vspace{0.1cm}
\begin{center}
\unitlength 1cm
\begin{picture}(10.,11.)
\put(-1.3,6.){\includegraphics{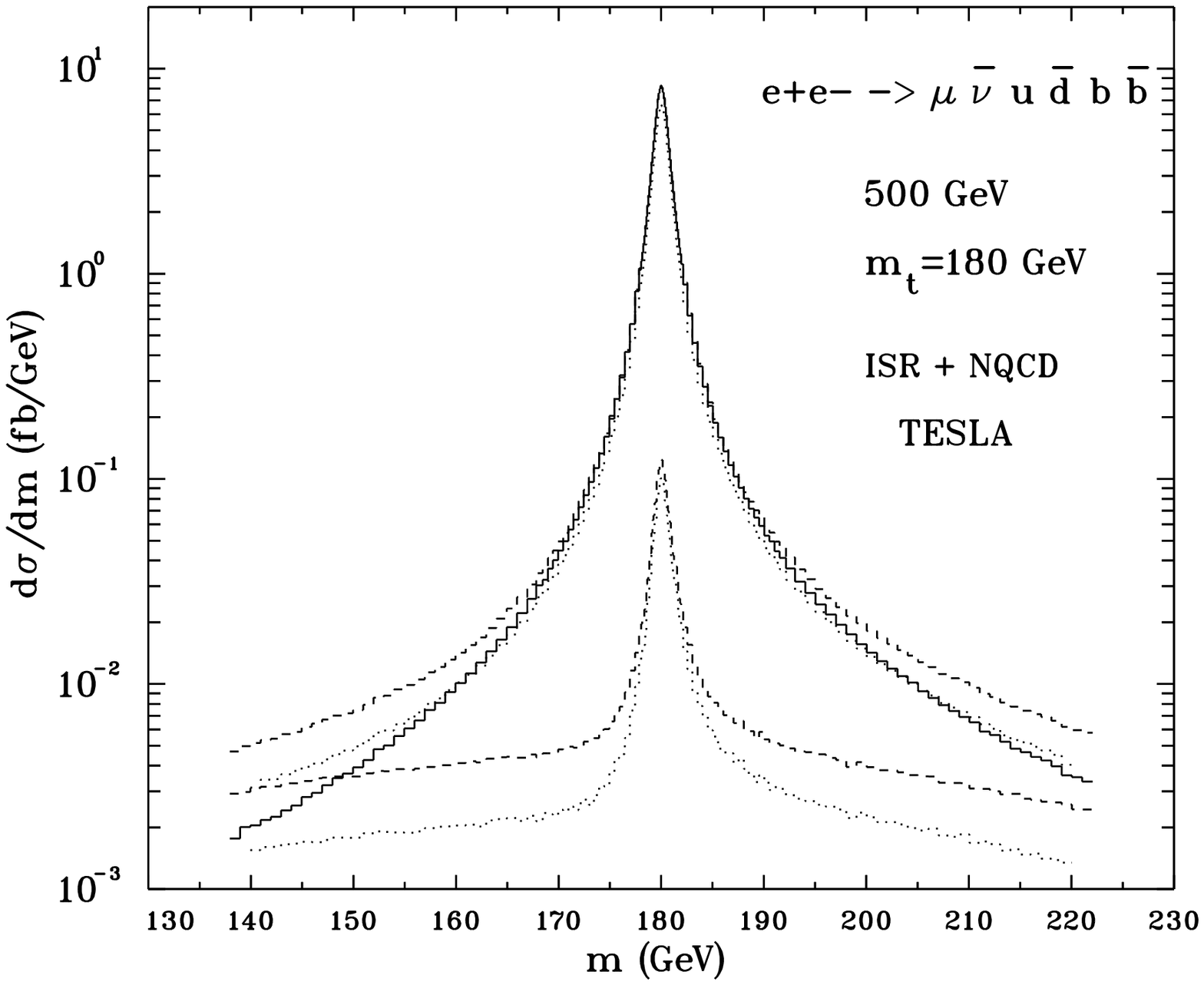}}
\put(-1.3,-1.3){\includegraphics{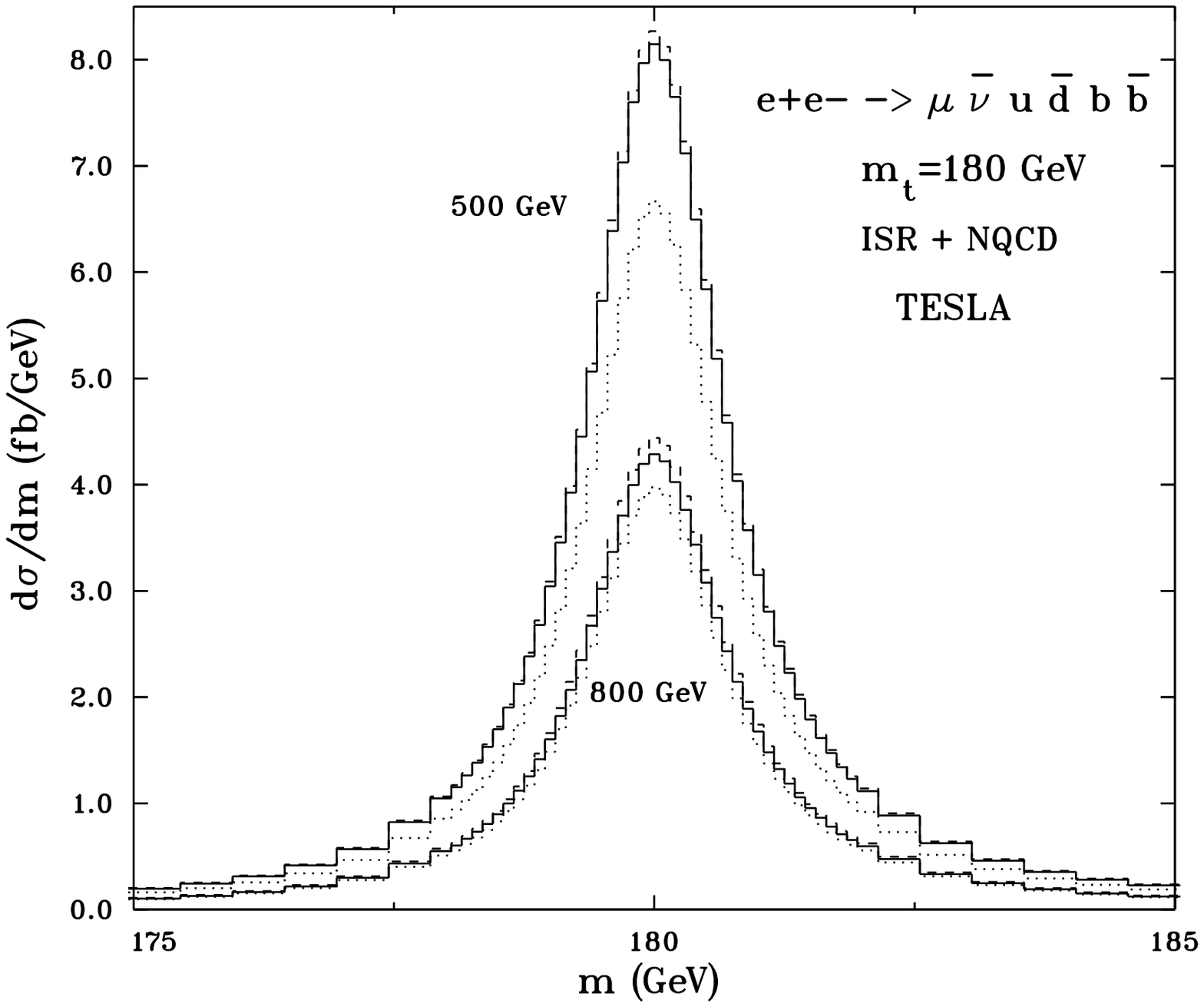}}
\end{picture}
\end{center}
\vspace{0.1cm}
\caption[]{Invariant mass distributions for top candidate. The nearest to the
nominal top mass between $u\bar db$ or $u\bar d\bar b$ invariant masses is
chosen event by event.
In the upper part the two lower curves represent the irreducible background
with and without cuts described in the text. All other lines represent
the contribution of $t\bar t$ off shell signal (solid), the full process
(dash) and the full process after cuts (dot). 
}
\label{f2}
\end{figure}

\begin{figure}[h]
\vspace{0.1cm}
\begin{center}
\unitlength 1cm
\begin{picture}(10.,6.5)
\put(-1.3,-1.3){\includegraphics{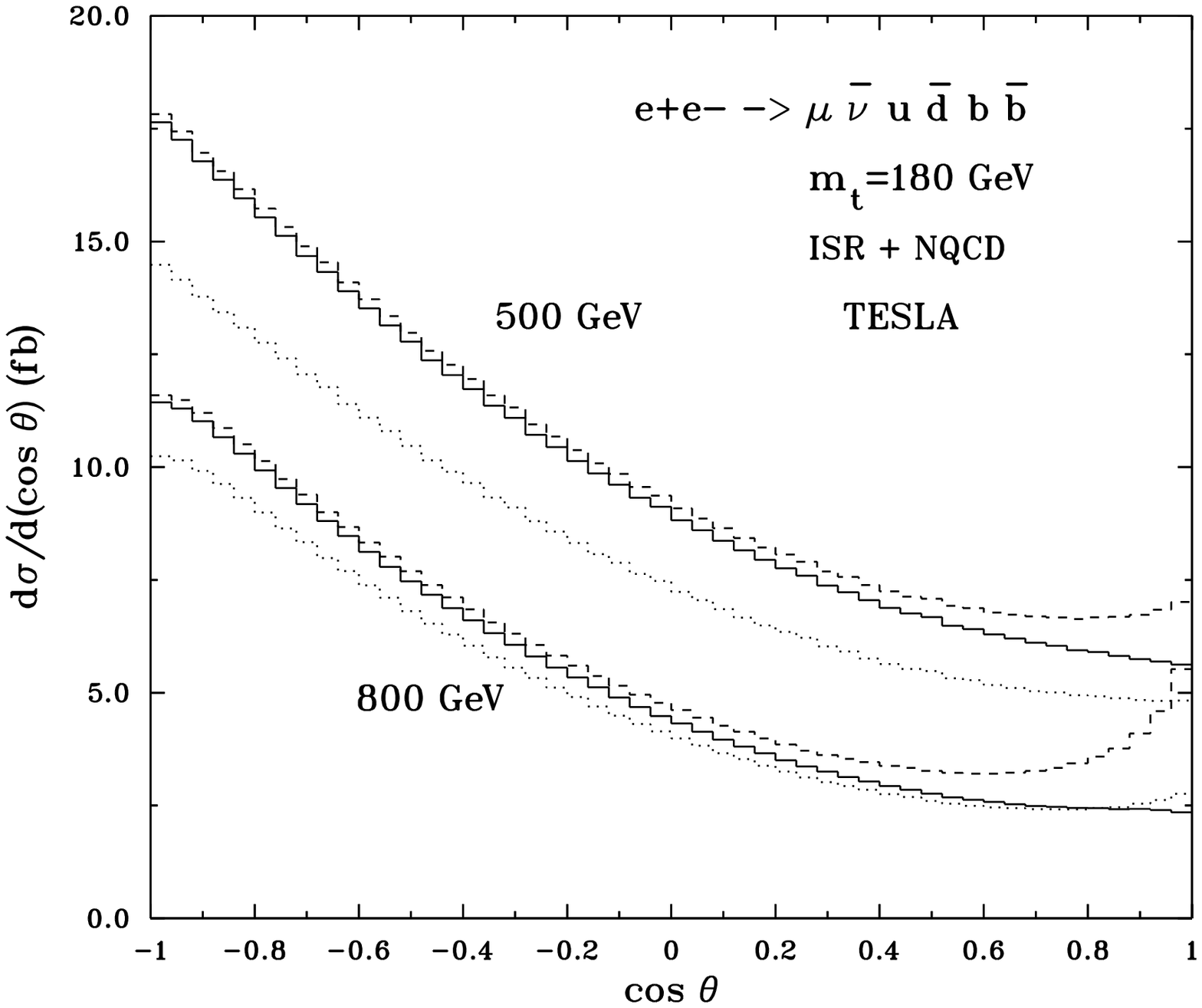}}
\end{picture}
\end{center}
\vspace{0.1cm}
\caption[]{Angular distribution for top candidate at $\sqrt{s}=500$ GeV (upper)
and $\sqrt{s}=800$ GeV(lower). 
The solid lines represent the contribution of $t\bar t$
off shell signal, the dash lines the full process  and the dot lines 
the full process after cuts. 
}
\label{f4}
\end{figure}

 Let us now  consider  the irreducible background
due to all diagrams contributing to a final state. From table~\ref{tabcstop} 
 one can conclude that the background  represents a 
correction of about 4\% to the signal.  It is however important to understand,
with the help of figs.~\ref{f2} and \ref{f4}, where  this difference manifests 
itself in the distributions. 
\vfill
In fig.~\ref{f2} it is reported 
the invariant mass distribution of the top candidate in the channel 
$ e^+e^-\ra \mu\bar\nu u \bar d b \bar b$. The differential cross sections
for  $ e^+e^-\ra e\bar\nu u \bar d b \bar b$ are practically identical. 
From table~\ref{tabcstop} one can easily see in fact
that the difference between the two channels ($\mu$ and e) is sizeable 
with respect to the backgrounds themselves, but it amounts to only a few 
permill of the signal. 

For a process of this kind, one should try to identify
the particles which might come from a  top decay, in order to measure their
invariant mass.
It is experimentally  difficult to reconstruct the
invariant mass of  the muon, its neutrino and  $\bar b$, as
 the neutrino momentum can only be deduced  from missing momentum, 
to which also ISR and BST contribute. 
It is probably better to look for the three quarks forming the top. 
We assume that there is b tagging, and we require that both b's are identified,
so that they are separated from the other two quarks.
One cannot  distinguish between a $b$ and a $\bar b$, therefore one
measures both the two invariant masses formed with one of the two b's and
the two other light quarks. Considering both values for every event, 
one obtains a  distribution with which  it will be possible to measure the 
mass of the top. Once this has been measured, one can refine the sample and 
look event by event for the nearest 
to the expected top mass between the two.
We refer to  this one as the mass of the top candidate, and we have plotted
its distribution. In fig.~\ref{f2} one can see  the differences among
 this distribution for the full process, the one due to the signal diagrams 
and that obtained with only  background diagrams. 
 This last distribution too  peaks at the top mass.
 This is of course the effect of  the many diagrams which are "single resonant",
in which one of the two top propagators  can go on mass shell.
 We have also computed the  same distribution after some cuts have been 
applied in order to try to  eliminate the background (dotted lines).
 The cuts we have imposed are:
\be
\rm   |m(ud)-m_W|<20\:GeV \qquad |m(b\bar b)-m_Z|>20\:GeV \label{cut1}
\ee
One can see that these cuts reduce in fact the background by about a factor two,
 but they do not affect its peak at the top mass. 

In the lower part of fig.~\ref{f2} we have reported on a linear scale and on 
the neighborhood 
of the peak the three curves relative
to full process, signal and full process after cuts, both at 500 GeV and 
at 800 GeV. 
From these curves one concludes that there does not seem to be 
any difference in the location of the maximum, but there are some appreciable
differences in the height between signal and total distributions.
The cuts (\ref{cut1}) are practically useless  in this region.

We have also studied the angular dependence of the top candidate. 
If one compares the dashed and full curves of fig.~\ref{f4}, one
notices that both at 500 GeV and at 800 GeV, there is a difference in the
angular distribution between signal and total calculations. This is
particularly relevant in the forward $e^+$ direction. 
The contribution of the irreducible background, may be reduced if
 one imposes  the cuts~(\ref{cut1}) and a cut
 on the mass of the top candidate $m_{tc}$: 
\be
|m_{tc}-m_t|<40\:GeV \label{cut2}
\ee  
As it can be seen from the dotted curves, there remains however a mild 
distortion of the total curve  with
respect to the signal one,  also after the cuts.

\section {WWZ and its background}

The main interest in WWZ  production lies in the possibility 
 to measure  gauge couplings. In particular 
the quartic gauge coupling and its possible deviations from SM will be 
studied at future $e^+e^-$ colliders. 
Several authors \cite{wwz,miya} 
have already analyzed three vector boson production and anomalous gauge
couplings. LHC measurements on VV+X (V=W,Z) final states will probably  reach a
better quartic gauge sensitivity than the WWZ measurement at 500 GeV $e^+e^-$
colliders.\cite{doba,miya},  nevertheless a careful six fermion study
of triple boson production is needed. In particular the
possible  background from $t \bar t$  production or top resonant diagrams
 may be dangerous and has to be analyzed in detail.
We consider in this section  WWZ signal and its irreducible 
backgrounds in  the  channels with four quarks and 
an isolated lepton. Among these, discriminating with b-tagging those 
final states not containing
$b$ quarks can help to reduce most of $t \bar t$ background.

The diagrams one has to deal with are similar to those 
of the preceding section. The total cross section is however lower
by an amount comparable to $t \bar t$ production as this type of diagram
is now no more resonant when there are no final b's.
The most important contribution 
comes from the 15 diagrams which correspond to WWZ production and decay.
 We will call them signal diagrams. The remaining ones can be divided in
 double, single and non resonant parts.
 
The results we will present  take into account ISR, BST, NQCD.

In table~\ref{cswwz500}  we present the cross sections for the full processes, 
those
computed taking into account signal diagrams  only, and those
computed via the \pd approximation. This
last result has  been obtained  taking the narrow width
approximation limit of the signal diagrams. 

\begin{table}[hbt]\centering
\begin{tabular}{|c|c|c|c|}
\hline
\rule [-0.25 cm]{0 cm}{0.75 cm}
process                         & WWZ NWA (fb) & WWZ signal (fb) & complete (fb)
\\
\hline \hline
\rule [-0.25 cm]{0 cm}{0.75 cm}
$\mu \bar \nu u\bar d c\bar c$  & 0.13836(2)   & 0.13464(2)      & 0.16218(9) \\
\cline{1-1} \cline{4-4}
\rule [-0.25 cm]{0 cm}{0.75 cm}
$e \bar \nu u\bar d c\bar c$    &            &             & 0.1783(2) \\
\hline
\rule [-0.25 cm]{0 cm}{0.75 cm}
$\mu \bar \nu u\bar d s\bar s$  & 0.17780(3) & 0.17303(3)  & 0.1803(1) \\
\cline{1-1} \cline{4-4}
\rule [-0.25 cm]{0 cm}{0.75 cm}
$e \bar \nu u\bar d s\bar s$    &            &             & 0.2117(2) \\
\hline
\rule [-0.25 cm]{0 cm}{0.75 cm}
$\mu \bar \nu u\bar d u\bar u$  & 0.12815(2) & 0.12469(2)  & 0.1512(1) \\
\cline{1-1} \cline{4-4}
\rule [-0.25 cm]{0 cm}{0.75 cm}
$e \bar \nu u\bar d u\bar u$    &            &             & 0.1758(3) \\
\hline
\rule [-0.25 cm]{0 cm}{0.75 cm}
$\mu \bar \nu u\bar d d\bar d$  & 0.16468(3) & 0.16025(3)  & 0.16733(9) \\
\cline{1-1} \cline{4-4}
\rule [-0.25 cm]{0 cm}{0.75 cm}
$e \bar \nu u\bar d d\bar d$    &            &             & 0.1941(1)  \\ 
\hline
\end{tabular}
\caption[]{Cross section for the processes $e^+e^-\ra l\bar \nu_l + 4$ light
quarks ($l=\mu,e$) at $\sqrt{s}=500$ GeV}
\label{cswwz500}
\end{table}

The difference between the full calculation and on shell (NWA) approximation is
indeed remarkable. Even between signal and NWA
there is a variation of some percent.  The background is much higher in a 
process with an up-type quark
pair than in the analogous one with down-type. For instance 
$\mu \bar \nu u\bar d c\bar c$  background is .0275fb while the 
$\mu \bar \nu u\bar d s\bar s$ is .0073fb. This corresponds to the fact  that,
for the set of cuts we are using, most background comes from diagrams with two
resonant W's and a $\gamma$ converting in a quark-antiquark pair. Diagrams
 with  one resonant W and the pair coming from Z decay would in fact produce 
an opposite behaviour.
\begin{figure}[h]
\vspace{0.1cm}
\begin{center}
\unitlength 1cm
\begin{picture}(10.,6.5)
\put(-1.3,-1.3){\includegraphics{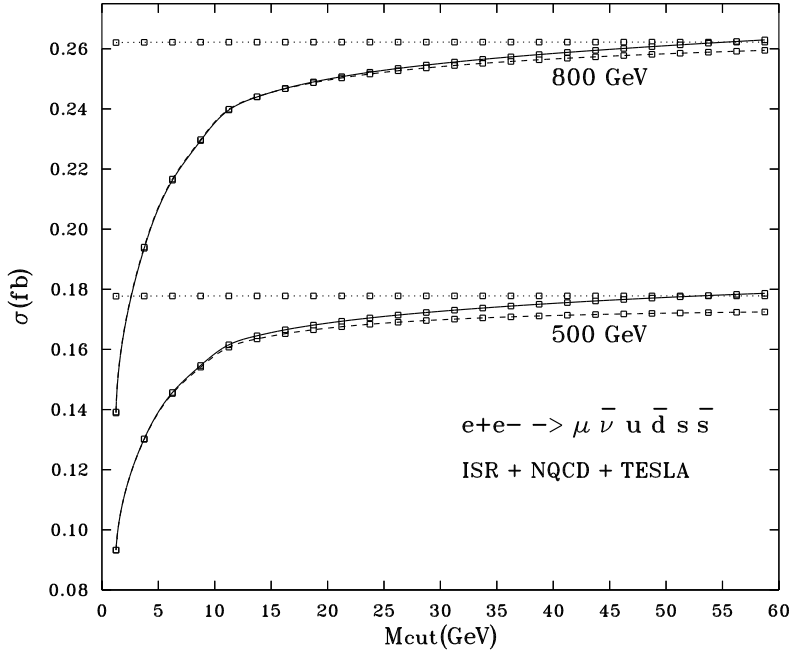}}
\end{picture}
\end{center}
\vspace{0.1cm}
\caption[]{Cross section for the process $e^+e^- \ar \mu \bar \nu_{\mu}
 u \bar d s \bar s$ as a function of $M_{cut}$ at $\sqrt{s}=500$ GeV (lower) and
 $\sqrt{s}=800$ GeV (upper). 
%It is required that two couples of quarks
%form two invariant masses $m_i$ ($i=1,2$) such that $|M_W - m_i| < M_{cut}$. 
Quarks are required to form two pairs whose invariant masses $m_i$ ($i=1,2$)
satisfy the conditions $|M_V -m_i| < M_{cut}$, $V=W,Z$.
The dot lines represent the cross section due to WWZ on shell, the dashed ones
the contribution of resonant WWZ diagrams only, the continuous the complete
cross section. The markers indicate the points effectively computed.
}
\label{f5}
\end{figure}
%\vfill
\begin{figure}[h]
\vspace{0.1cm}
\begin{center}
\unitlength 1cm
\begin{picture}(10.,6.5)
\put(-1.3,-1.3){\includegraphics{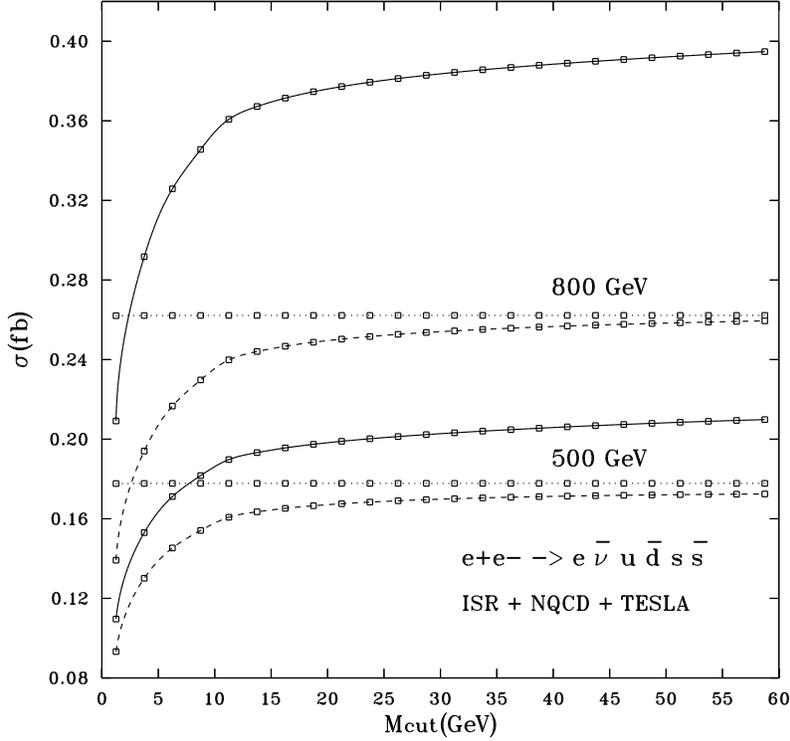}}
\end{picture}
\end{center}
\vspace{0.1cm}
\caption[]{Cross section for the process $e^+e^- \ar e^-\bar \nu_e u
\bar d s \bar s$ at $\sqrt{s}=500$ GeV (lower) and $\sqrt{s}=800$ GeV (upper)
as a function of $M_{cut}$.
The definition of $M_{cut}$ and the meaning of the different lines and of the
markers is the same as in fig~\ref{f5}.}
\label{f6}
\end{figure}
%\vfill
\begin{figure}[h]
\vspace{0.1cm}
\begin{center}
\unitlength 1cm
\begin{picture}(10.,6.5)
\put(-1.3,-1.3){\includegraphics{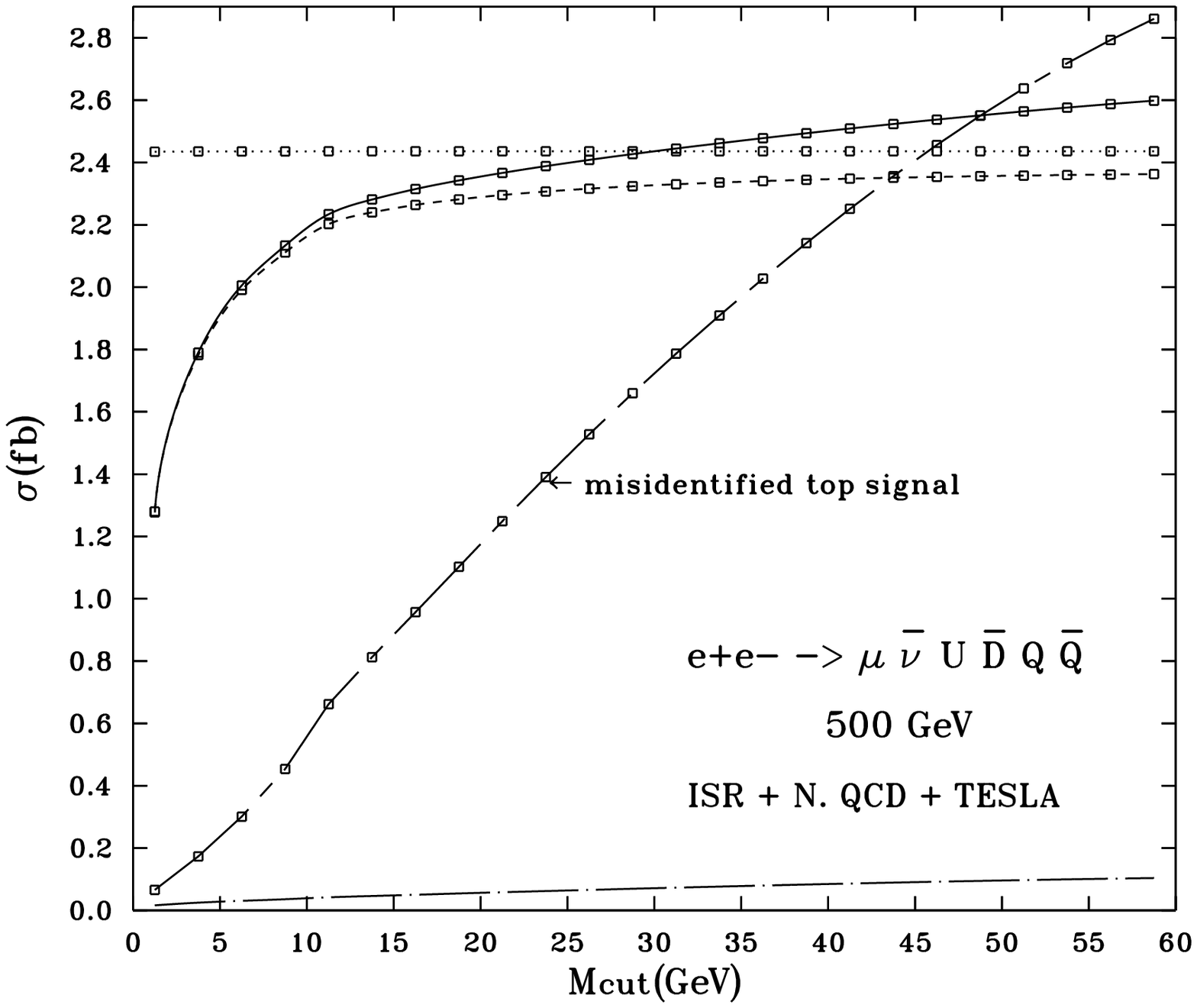}}
\end{picture}
\end{center}
\vspace{0.1cm}
\caption[]
{cross section for $e^+e^-\ra \mu \nu U D Q\ol Q $ as a function of 
$M_{cut}$. We sum over all processes with one muon and no b quarks.
The chaindash (chaindot) line represents the background due to top top signal
(background) with both misidentified b's. 
The definition of $M_{cut}$ and the meaning 
of the different lines and of the markers is the same as in fig~\ref{f5}.}
\label{f8}
\end{figure}
%\vfill

In presence of
such big irreducible background, it is necessary to introduce  cuts 
in order to try to isolate WWZ production.
We have implemented in figs.~\ref{f5},\ref{f6} and \ref{f8} 
different cuts on invariant masses of the pairs of 
particles which should come from vector bosons, and computed the cross sections
with the full set of diagrams (continuous line), 
with only the fifteen signal diagrams (dashed),
and with \pd or narrow width approximation (dotted).
 Obviously  these last cross sections are  not sensible
to such kind of cuts.
We consider at first some cuts which act only on the quarks, as the 
neutrino momentum is not directly measurable. 
We accept an event if out of the three couples of pairs of quarks, one at least
has a pair  whose invariant mass $m_1$ satisfies $|M_W -m_1| < M_{cut}$
and the other pair's mass $m_2$ satisfies $|M_Z -m_2| < M_{cut}$.
In  figs.~\ref{f5}-~\ref{f6} are reported the corresponding cross sections 
as a function of $M_{cut}$.
 Both figures show that the signal contribution  to the cross section 
is lower than the NWA even for very loose cuts or for the cross section 
without cuts (see table~\ref{cswwz500}). 
The curves of fig.~\ref{f6}  show that for the electron case the difference 
between signal and total process is extremely relevant. It grows with the 
energy and
the cuts we have imposed can greatly reduce the difference but not suppress it.
For the muon case an $M_{cut}$ of about 10 GeV is on the contrary sufficient
to make total and signal cross section practically coincide. The loss in
event number is however of the order of ten percent.
In order  to further suppress the background in the electron case, 
we have  also  imposed a cut
on the invariant mass $M_{rec}$ formed with the four momentum of the 
electron and the reconstructed  neutrino one. In such a case  we attribute  
all missing three-momentum $\bar p_{mis}$ to the neutrino and take its energy 
to be equal to $|\bar p_{mis}|$.
We have indeed verified that even a very loose cut as  
$|M_W -M_{rec}| < 60$ GeV reduces significantly the difference between signal
and total cross sections. 
On the other hand, 
if the cut has to be so stringent as to reduce the difference between signal 
and total cross section to the order of the percent,  i.e. 
$|M_W -M_{rec}| < 15$ GeV, one looses  about one third 
of the event number, as compared to the NWA. This conclusion is important
in view of the fact that such WWZ processes have a low cross section:
at 500 GeV $\rm \sigma(WWZ)$ is of the order of 40fb, which corresponds
to a total of 2000 events  for an integrated luminosity of 50 $\rm
fb^{-1}$.

The reactions  studied in figs.~\ref{f5}-\ref{f6} 
 cannot be directly measured, as the different
 quark flavours cannot be disentangled experimentally.
 In fig.~\ref{f8} we examine the more interesting physical case
in which we sum over all reactions involving one isolated muon. 
In the plot it is reported, both for WWZ diagrams  and for the complete
calculation, the sum of the cross sections for 
\[
  e^+e^-\ra \mu^- \bar \nu_\mu u \bar d s \bar s \quad
  e^+e^-\ra \mu^- \bar \nu_\mu u \bar d c \bar c  \quad
  e^+e^-\ra \mu^- \bar \nu_\mu u \bar d u \bar u \quad
  e^+e^-\ra \mu^- \bar \nu_\mu u \bar d d \bar d  
\]
multiplied by a factor 4. This accounts for the reactions in which one has 
$\mu^+  \nu_\mu \bar u d$ 
instead of $\mu^- \bar \nu_\mu u \bar d$ and for those in which one 
has $c \bar s$ (or $\bar c s$) instead of $u \bar d$ (or $\bar u  d$).
 The dashed and continuous lines of fig.~\ref{f8} give therefore
the total  cross section as a function of $M_{cut}$ for 
all processes with one  muon, four quarks  and no b's in the final state.

 In order to reduce  the enormous  background 
from $t \bar t$ production and decay, b-tagging will be used. 
With it, one can exclude all events with at least one tagged b. 
 With actual b-tagging techniques, this however leads to a reduction of
the signal without completely suppressing the background.
In fact, if there is a high probability $P_{c\ra b}$
that a $c$ be misidentified as a $b$, one has to
multiply by the appropriate reduction factors $1-P_{c\ra b}$,
$(1-P_{c\ra b})^2$, $(1-P_{c\ra b})^3$
the contributions with one, two or three $c$'s
to the the above sums.
This would give a decrease of the signal curves of about 25\% for 
$P_{c\ra b}=.3$. Moreover if there is a finite probability 
$P_{b\ra \slaar b}=1-P_{b\ra b}$ that a b
may not be recognized as such,  events  with two b's
can be misidentified  and give a residual
 background to WWZ physics.  The chain dash and chain dot curves of 
fig.~\ref{f8} correspond  to such a background as a function of $M_{cut}$.
We have assumed $P_{b\ra \slaar b}=.2$ and summed over the processes 
\[
  e^+e^-\ra \mu^- \bar \nu_\mu u \bar d b \bar b \qquad
  e^+e^-\ra \mu^- \bar \nu_\mu c \bar s b \bar b  \qquad
  e^+e^-\ra \mu^+      \nu_\mu \bar u d b \bar b \qquad
  e^+e^-\ra \mu^+      \nu_\mu \bar c s b \bar b.
\]
From these curves, one can conclude 
that even an imperfect b tagging is of considerable help in strongly 
reducing the great number of these events. It has to be remarked that
this background depends strongly on the applied
$M_{cut}$. If one adopts the strategy of applying a severe $M_{cut}$ 
of the order of 10 GeV, it is reduced to about 1/6 of WWZ signal.
If on the other hand a milder $M_{cut}$ is used or if $P_{b\ra\slaar b}$ is
greater than what we used, it may become comparable to the signal itself.

\section{Intermediate mass Higgs.}
\begin{figure}[h]
\vspace{0.1cm}
\begin{center}
\unitlength 1cm
\begin{picture}(10.,10.)
\put(-1.3,-2.){\includegraphics{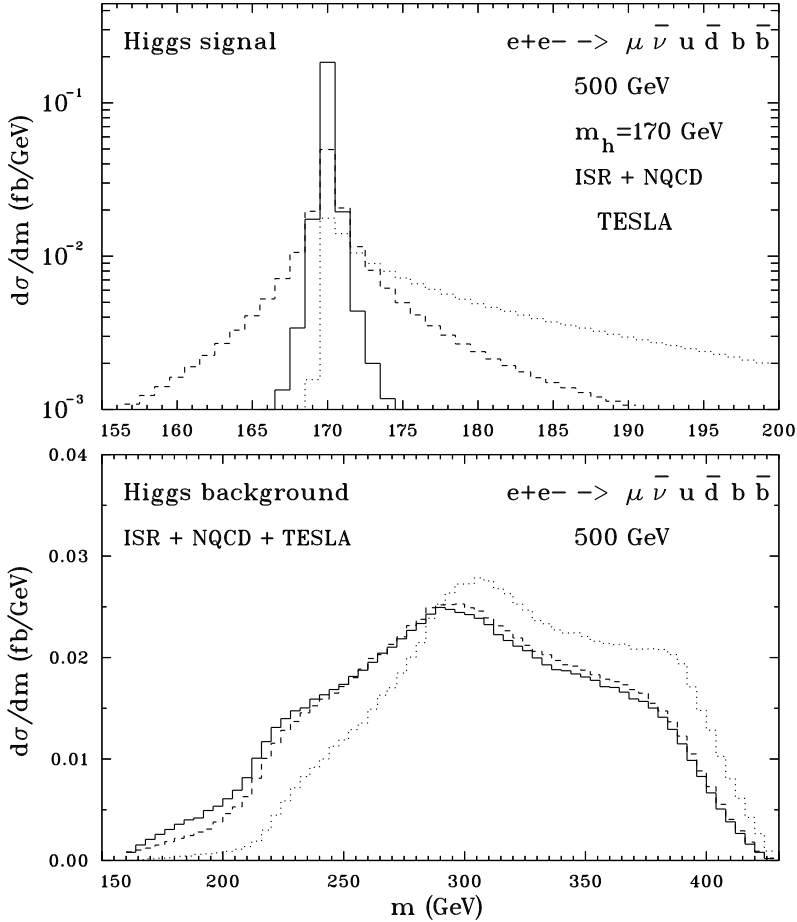}}
\end{picture}
\end{center}
\vspace{0.1cm}
\caption[]{Invariant mass distributions for Higgs signal (upper) and
background (lower). The continuous line corresponds to ($\mu \bar \nu
u \bar d$)  mass, the dashed  to the {\it reconstructed} and the dot to the 
{\it missing} one.}
\label{h1}
\end{figure}

\begin{figure}[h]
\vspace{0.1cm}
\begin{center}
\unitlength 1cm
\begin{picture}(10.,10.)
\put(-1.3,-2.){\includegraphics{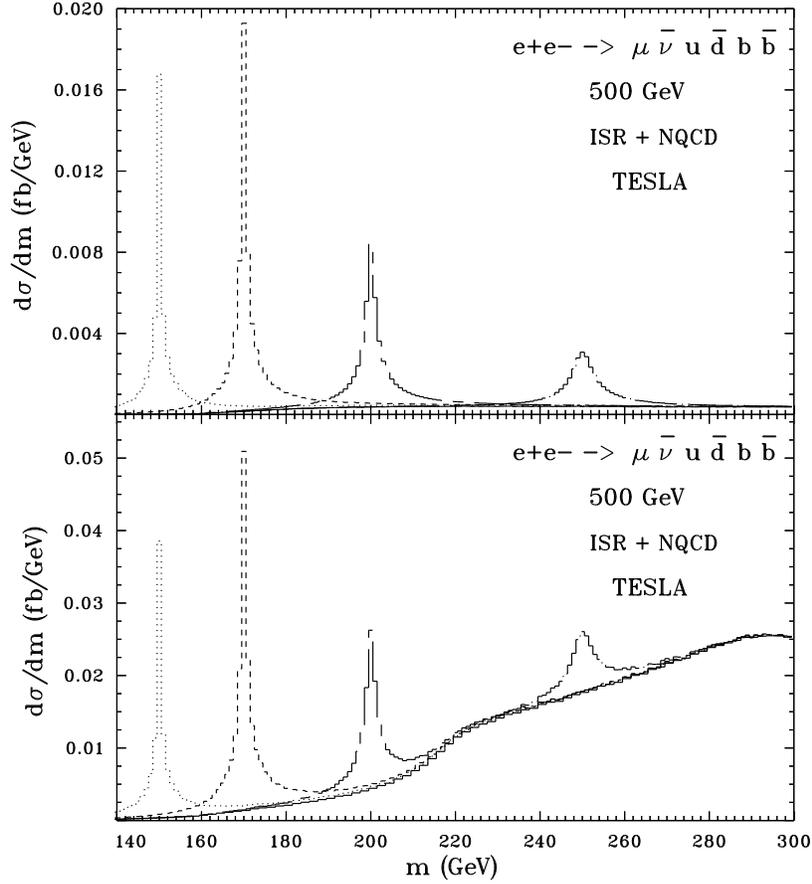}}
\end{picture}
\end{center}
\vspace{0.1cm}
\caption[]{{\it Reconstructed} mass distributions with {\it normal} cuts 
(lower) 
and {\it normal}  cuts + $|M-m_{top}|>$40 GeV ( $M=m(bu\bar d)$ and 
$m(\bar bu\bar d)$ ) (upper). The continuous line represents the total 
background. The others correspond to the total cross sections for 
(from left to right) $m_h$= 150, 170, 200, 250 GeV.
 }
\label{h2}
\end{figure}

\begin{figure}[h]
\vspace{0.1cm}
\begin{center}
\unitlength 1cm
\begin{picture}(10.,10.)
\put(-1.3,-2.){\includegraphics{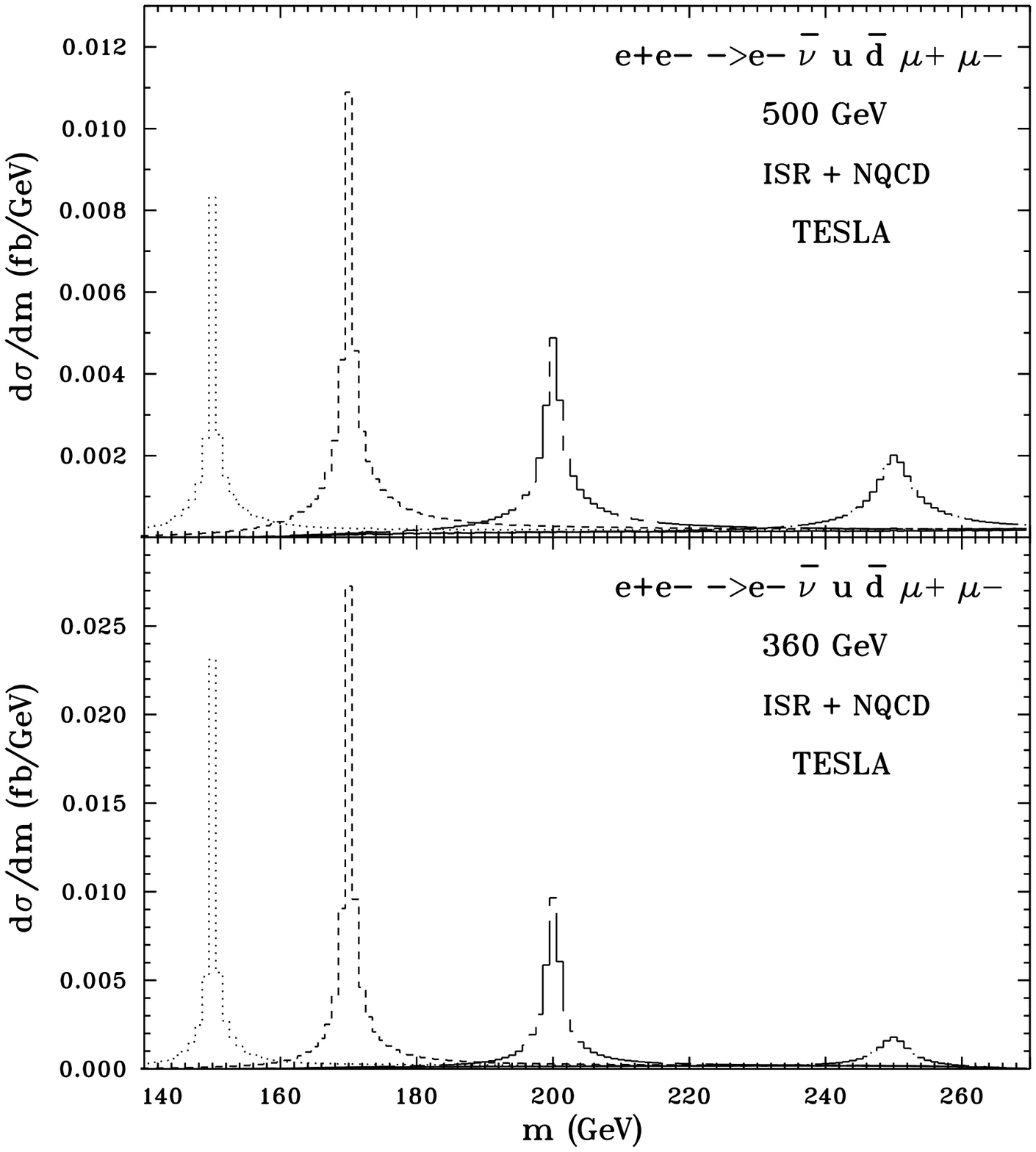}}
\end{picture}
\end{center}
\vspace{0.1cm}
\caption[]{{\it Reconstructed} mass distributions with {\it normal} cuts . 
The continuous line represents the total 
background. The others correspond to the total cross sections for 
(from left to right) $m_h$= 150, 170, 200, 250 GeV.
}
\label{h3}
\end{figure}

\begin{figure}[h]
\vspace{0.1cm}
\begin{center}
\unitlength 1cm
\begin{picture}(10.,10.)
\put(-1.3,-2.){\includegraphics{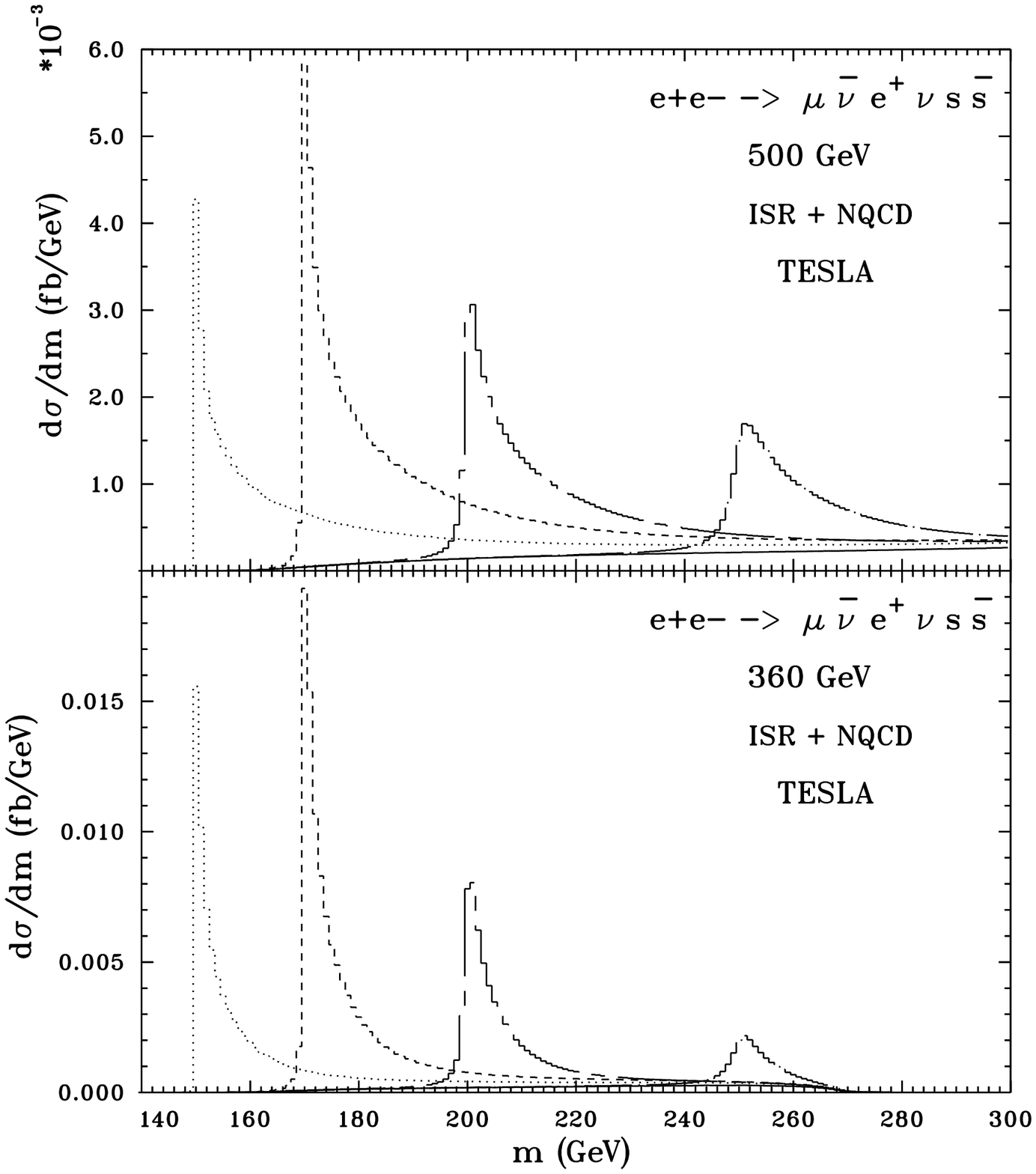}}
\end{picture}
\end{center}
\vspace{0.1cm}
\caption[]{{\it Missing} mass distributions with {\it normal} cuts . 
The continuous line represents the total 
background. The others correspond to the total cross sections for 
(from left to right) $m_h$= 150, 170, 200, 250 GeV.
}
\label{h4}
\end{figure}
We will consider in this section the scenario in which the Higgs mass is in the
intermediate range. As it is well known a light Higgs decays predominantly
to \bb pairs. When the mass of the Higgs is of the order of 140 GeV,
\bb and $W W^*$ decays become comparable. After the threshold for two
W's production, $W^+ W^-$ decay becomes the dominant one, its branching being
almost one at threshold and of the order of .7 for higher  masses, where  
two Z's decay is kinematically allowed. 
As far as production of the Higgs is concerned, the dominant process
at \epem is HZ production (Higgs bremstrahlung) up to a c.m. energy of 500 
GeV, where the WW fusion process has about the same magnitude as the 
previous one. At higher energies this last process dominates.

Most Higgs events will in effect appear as six fermion
events for a mass above 140 GeV. Those coming from WW fusion will be
characterized by two neutrinos and four fermions coming from the
chain decay $H\ar W^+W^-\ar 4f$. If one of the W's decays leptonically, there
will be at least three neutrinos around, and it will be difficult to reconstruct
the Higgs invariant mass. Therefore WW fusion events will  be mainly studied
in 4q + 2$\nu$'s final state. This implies that six fermion background
to these events will come  from  NC, MIX and CC  processes and it will be rather
complicated to deal with. 

 We limit our present analysis to CC processes. We  examine therefore in some 
detail Higgs bremstrahlung for an intermediate mass Higgs in the following
final states:
\[
 1)\; l\: \nu_l\: +\: 4\: q's, \hsk\hsk 
2)\; l\: \nu_l\: +\: l'\: \bar l'\: +\: 2\: q's,\hsk\hsk
3)\; l\: \nu_l\: +\: l'\: \nu_{l'}\: +\: 2\: q's.
\]
Typical examples are respectively $e^+ e^- \ar \mu \bar \nu u \bar d b \bar b$,
$e^+ e^- \ar e \bar \nu u \bar d \mu^+  \mu^- $,  
$e^+ e^- \ar \mu \bar \nu_\mu e^+  \nu_e s \bar s$.
Considering the  branching of the W's and of the Z, one can deduce that
the percentage of all processes of type 1) with respect to the full HZ signal
is about 31\%. The others are approximately 4.4\% for 2) and 5.2\% for 3).
 These values  have been obtained including  also the $\tau$'s. In  case 3)
we have not taken into account the decays in which $l$ and $l'$ have 
the same  flavour. Let us recall that at a center of mass energy of 500 GeV
one has a cross section for $e^+e^- \ar HZ$ of approximately 40~fb for an Higgs
mass of the order of 200 GeV. For an integrated luminosity of 50~fb$^{-1}$ per 
year, this means around 2000 HZ events and about 600 events for case 1), 
100 each  for  2) and 3). 
The other type of signal events which can be studied with CC cross sections
is   $l\: \nu_l$ + $l'$ $\nu_{l'}$ + $l''\; \bar l''$ but it amounts only to .7\%
of the ZH events and will not be considered. All other events with two 
$\nu$'s of the same flavour in the final state and/or with both W's decaying
hadronically cannot be discussed without the full NC, MIX and CC
contributions to the irreducible background.

The isolated lepton (e.g. a $\mu^-$ without the corresponding $\mu^+$)
characteristic of events 1), 2), 3) can be considered an experimental signature
that in the process two W's have been produced. It is therefore extremely useful
in reducing the background. On the other hand 
the neutrino makes  the reconstruction of the Higgs mass more difficult, as 
missing energy and momentum are also  due to ISR and BST.

The main six fermion background to events of type 1) comes from $t \bar t$.
With  b-tagging it  can be greatly reduced  when the 4 q's are light ones.
In this case it is however more difficult to find out which pair of quarks 
pertains to the Z and which enters in H mass reconstruction. The signal events
of type 1) with 2 b's are about 6.8\% of the total HZ signal. 

In order to properly understand  $t \bar t$ background, we take as a case
study  the reaction $e^+ e^- \ar \mu \bar \nu u \bar d b \bar b$. 
The most relevant distribution to analyze in Higgs physics  is the invariant
mass of the particles which in the signal diagram decay  from the Higgs.
This distribution is however not directly measurable when at least one of the
W's decays leptonically. 
We will  therefore  consider two other distributions. 
We will call  {\it reconstructed}  the distribution
in which all  missing 3-momentum is attributed to the neutrino,
and its energy is taken to be equal to its modulus. 
We will instead name {\it missing} the distribution of the invariant mass
of the 4-momentum recoiling from the Z-decay particles 
($P_{tot}-P_b-P_{\bar b}$ in the case at hand). 
In fig.~\ref{h1} are reported the above distributions both for the signal
($m_h=170$ GeV) and for the background. One notices an expected broadening
of the peak for the {\it reconstructed} distribution. The {\it missing}
distribution produces in addition a typical distortion of the signal and a 
shift of the background toward higher masses. For such a reason we will examine
in the following {\it reconstructed} distributions, apart from
processes of type 3) above, where only the {\it missing} one is 
measurable, due to the presence of two neutrinos whose 3-momentum
cannot be separately reconstructed. Here and in the following
we apply  the request that the invariant
mass of the two particles decaying from the Z and the two quarks eventually
decaying from the W, be within 20 GeV from the Z and the W mass respectively.
We refer to these cuts, 
in addition to those described in the introduction, as {\it normal} cuts.
From fig.~\ref{h2} (lower) one can realize that such cuts are already rather
effective in reducing  $t\bar t$ background, expecially for low $m_h$ values.
One can try to reduce it further with the additional cuts 
$|m(bu\bar d)-m_{top}|>$40 GeV and $|m(\bar bu\bar d)-m_{top}|>$40 GeV. The 
upper part of fig.~\ref{h2} shows that in such a case the background becomes
completely negligible, but at the price of reducing also the signal by an
approximate factor 3.

In order to analyze intermediate Higgs production in the final states 2) and 3)
we consider the processes $e^+ e^- \ar e \bar \nu u \bar d \mu^+  \mu^-$ and
$e^+ e^- \ar \mu \bar \nu_\mu e^+ \nu_e s \bar s$ respectively.
For the first of the two, fig~\ref{h3} shows that with {\it normal}
 cuts and for the
{\it reconstructed} mass the whole contribution of the hundreds of background 
diagrams is practically almost irrelevant both at 360 and 500 GeV center of 
mass energy. For most studies a reasonable 
approximation in such a case would be to consider only the off-shell six 
fermion signal diagram $e^+e^-\ar H^*Z^* \ar W^{+*}W^{-*}Z^*\ar
 e \bar \nu u \bar d \mu^+ \mu^-$. A similar conclusion  (see fig.~\ref{h4})
applies also to 
$e^+ e^- \ar \mu \bar \nu_\mu e^+ \nu_e s \bar s$ at 360 GeV.
At 500 GeV the background becomes somewhat more important and the tail 
of the  {\it missing} distribution can be exactly reproduced only with the full 
calculation.

\section{Conclusions}
Six fermion processes  will become relevant at future $e^+e^-$ linear colliders.

With the help of the helicity amplitude method of 
ref.~\cite{phact}\cite{method}, we have built up  a program (SIXPHACT) to
compute all complete tree level charged current cross sections.
The program can also compute any differential cross section (distribution)
with high precision in reasonable computer time. 
 Complete six fermion calculations are in fact particularly useful 
when studying distributions, where one can understand the differences from 
the usual \pd approximation and analyze for instance  which cuts have to be 
imposed to enhance signals with respect to irreducible backgrounds.

We have given some examples  of phenomenological studies relevant
to top, WWZ and Higgs physics.
We have in particular found that single resonant background contributions are
difficult to get rid of when studying invariant mass or angular 
top distributions. For WWZ it seems that \pd approximation is not viable,
and that to get rid of the irreducible background with the cuts
we have tried, one looses about 10\%
of events with a muon in the final state and more than 30\% for an electron 
at 500 GeV.
The enormous amount of background coming from $t \bar t$ production 
is under control when  appropriate b-tagging and cuts are applied. 
For what concerns intermediate Higgs physics, 
we have restricted our analysis to final states with at least an isolated
lepton which most probably comes from a W decay. The most important contribution
to the signal comes from final states with four quarks. These events suffer
from  $t \bar t$ background which we have analyzed in detail
and found to be greatly reduced by a simple cut on the invariant mass of the 
quarks decaying from Z. Processes of the type 
$ l\: \nu_l\: +\: l'\: \bar l'\: +\: 2\: q's$ have a harmless irreducible 
background
with a cut on $ l\: \bar l$ invariant mass. 
Finally processes of the type $ l\: \nu_l\: +\: l'\: \nu_l'\: +\: 2\: q's$
can also give an important  contribution to Higgs studies if their {\it missing}
mass distribution is carefully analyzed.

\vfill\eject
\end{document}